\documentclass[pra,twocolumn,showpacs,amsmath,amssymb,floatfix]{revtex4-1}
\usepackage{graphicx}

\newcommand{\be}{\begin{eqnarray}}
\newcommand{\ee}{\end{eqnarray}}

\newcommand{\bfq}{{\bf q}}

\newcommand{\bfk}{{\bf k}}

\newcommand{\tlf}{\tilde{f}}

\newcommand{\wbe}{\begin{widetext}}
\newcommand{\wee}{\end{widetext}}
\newcommand{\oncite}{\onlinecite}

\begin{document}

\title{Quantum fluctuations and condensate fraction during the time-of-flight expansion}

\author{Shiang Fang$^{1}$, Ray-Kuang Lee$^{1,2}$ and Daw-Wei Wang$^{1,3}$}

\affiliation{
$^{1}$ Physics Department, National Tsing-Hua University, Hsinchu 300, 
Taiwan
\\
$^{2}$ Institute of Photonics Technologies, National Tsing-Hua University,
Hsinchu 300, Taiwan
\\
$^{3}$ Physics Division, National Center for Theoretical Sciences,
Hsinchu 300, Taiwan}

\date{\today}

\begin{abstract}
The quantum fluctuation effects in the time-of-flight (TOF) experiment for a 
condensate released from an optical lattice potential is studied within the truncated Wigner approximation. By investigating both the spatial and momentum density distributions, we find that the condensate fraction decreases monotonically in time and hence cannot be measured in the standard TOF image. We then propose a semi-quantitative analysis for such dynamical quantum depletion process. Our study shows a universal algebraic decay of the true condensate fraction, and have a very good agreement with numerical results. 
We also discuss possible methods to determine the condensate fraction inside the optical lattice, and its implication to the TOF experiments in higher dimensional systems.
\end{abstract}


\maketitle
Experimental realization of the superfluid to Mott insulator
transition in ultracold bosonic atoms 
opens a new horizon of strongly correlated physics [\oncite{SF_MI,review}]. Among various kinds of experimental techniques, 
time-of-flight (TOF) experiment is the most important tool to observe
possible quantum orders and related correlations [\oncite{Ian,Ho,Nature_fermion_pair_MIT}]. It is generally believed that the TOF image can be interpreted as the momentum distribution of the atom cloud, if (i) the final size of the atom cloud is much larger than the initial one and (ii) the interaction effect during the expansion is negligible. As a result, the sharp interference peaks near the zero and Bragg wavevectors can be interpreted as a signature of a Bose-Einstein condensation inside the optical lattice. 

Recently, Gerbier {\it et al.} [\oncite{QMC_Troyer}] show that the condition (i) is usually not satisfied, because the required time scale is much larger than the feasible value in the present experiments. As for the condition (ii), it is argued that the interaction effects are negligible (or at most in the meanfield level) due to the sudden drop of the particle density [\oncite{QMC_Troyer,duan,scaling}]. Different from the condensate in a single trapping potential, however, particles loaded in an optical lattice contain a much stronger interaction energy (and hence quantum fluctuation). In principle, the interaction effect can be removed by using Feshbach resonance, but such a far-from-equilibrium quantum dynamical problem itself is still very interesting and worthy of further investigation. It may also provide valuable insights for studying other quantum orders through the TOF experiment.
\begin{figure}
\includegraphics[width=8.2cm]{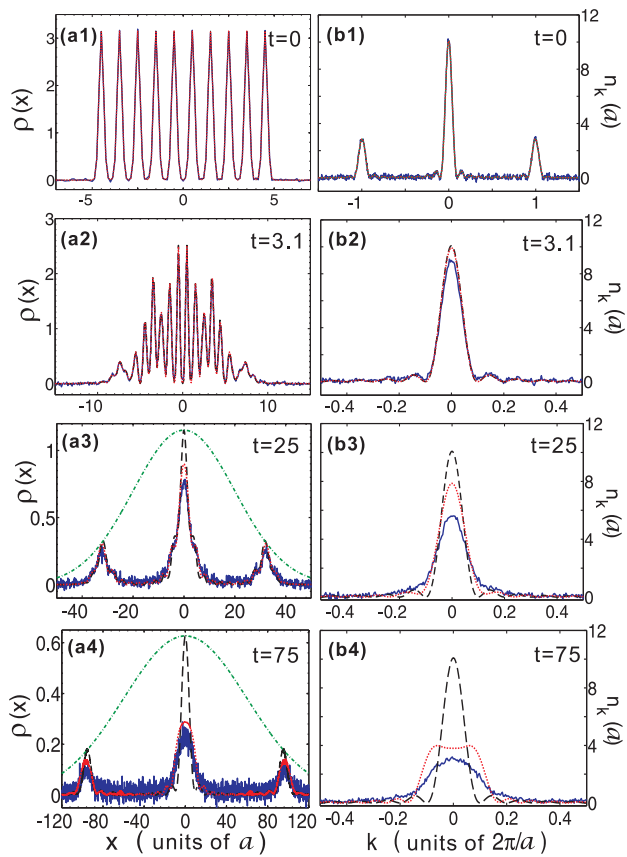}
\caption{(Color online) Time evolution of the spatial (a1)-(a4) and 
momentum (b1)-(b4) distributions for an interacting condensate released from a 1D optical lattice. The expansion time ($t$) is in the unit of $\hbar/E_R$.
The dashed black, dotted red, and solid blue lines are the non-interacting, GPE and TWA results, respectively. The dash-dotted green lines in (a3) and (a4) are proportional to the square of the expanding Wannier function. Other parameters are described in the text.
}
\label{main}
\end{figure}

In this paper, we numerically and analytically investigate the quantum dynamics of the TOF experiment for an interacting condensate released from a one-dimensional (1D) optical lattice potential. The quantum fluctuations are included within the truncated Wigner approximation (TWA). The common adapted empirical definition of the condensate fraction, measured from the bi-modal fitting of a spatial distribution, is found to saturate at a value substantially smaller than the free expansion result. On the other hand, the true condensate fraction, which can be unambiguously determined in the momentum space at any moment, decreases monotonically in time due to the dynamical quantum depletion. We find that the decaying condensate fraction can be fitted very well by a simple algebraic function with a universal exponent, showing an interaction induced quantum decoherence effects. We also show the 2D results and find a similar quantum depletion phenomenon. Our results provide a deeper understanding about such a far-from-equilibrium quantum dynamical process and can be applied to studying other quantum orders.

We consider $^{87}$Rb atoms loaded in a 1D optical lattice with the scattering length, $a_s=5.45$ nm, the lattice constant, $a=0.5$ $\mu$m, and the lattice strength, $V_0=10$ $E_R$, where $E_R=2\pi \hbar\times 2.3$ kHz is the recoil energy. The transverse confinement radius is $R=0.11$ $\mu$m, the on-site interaction energy is $U=0.2 E_R$. When $t>0$, the lattice potential is turned off and atoms start to expand with the same transverse confinement (with an effective 1D interaction strength, $g_1=\frac{4\pi\hbar^2 a_s}{2\pi R^2 m}$, where $m$ is the atomic mass). We use three approximations for numerical studies: non-interacting expansion, Gross-Pitaevskii equation (GPE), and the truncated Wigner approximation (TWA), respectively.
Note that the TWA can be shown to be {\it exact} in the short-time limit as well as in the weak interaction limit [\oncite{TWA_book,TWA_early_paper,ap1}], and therefore is very appropriate for the study of TOF dynamics. In all of our following calculations, we apply the same initial condition, a pure condensate state with ten particles in ten lattice site, for simplicity. Application of our results to different initial states (say Mott state or with finite thermal/quantum depletion) is feasible, but is not our interest in this paper. Finally, we discretize the optical lattice potential by eight mini-cells in space, and calculate the expansion dynamics of the whole system up to 2048 mini-cells. In other words, for the initial condensate of ten lattice sites, the atom cloud is allowed to expand at most 25 times larger than its initial size within 100 $\hbar/E_R\sim 7$ ms. 

In Fig. \ref{main}(a1)-(a4), we show the calculated spatial density distributions at different expansion time. We note that both the GPE (dotted red lines) and the TWA (solid blue lines) results show a strongly broadened central peak compared to the non-interacting result (dashed black lines), while the difference becomes much weaker in the two Bragg peaks. To quantify such an interaction effect, we define the following renormalized peak ratio, $\eta$:
\be
\eta(t)\equiv \frac{\rho(0,t)}{\rho(x(2\pi /a),t)}
\times\frac{|\tilde{w}(2\pi/a)|^2}{|\tilde{w}(0)|^2},
\label{eta}
\ee
where $\tilde{w}(k)$ is the Fourier transform of the Wannier function at the wavevector $k$,$x(k)=\hbar tk/m$ ,  and $\rho(x,t)$ is the particle density at position $x$ at time $t$. We note that $\eta$ is exactly equal to one for a free expansion. This can be shown from the free expansion result: $\rho(x(\bf k),t)\propto |\tilde{w}(\bfk)|^2{\cal S}(\bfk)$ [\oncite{QMC_Troyer}], where the structure function, ${\cal S}(\bfk)$, is a periodic function with the period $2\pi/a$, i.e., ${\cal S}(0)={\cal S}(2\pi/a)$. Nevertheless, when the interaction is considered, our TWA result shows that $\eta$ can be reduced to about 0.65 at $t=75\hbar/E_R$ for the simulations in Fig. \ref{main}. In fact, similar experimental result has been observed in Ref. [\oncite{peak_ratio}], where the population ratio of the first Bragg peak to the central peak is 35\% larger than the ratio of a free expansion (i.e., $\eta(t)=\frac{1}{1.35}\sim 0.74$) at $t=29.5$ ms. A systematic study on $\eta(t)$ should provide more information about the meanfield interaction effects during the TOF experiment.

\begin{figure}
\includegraphics[width=7cm]{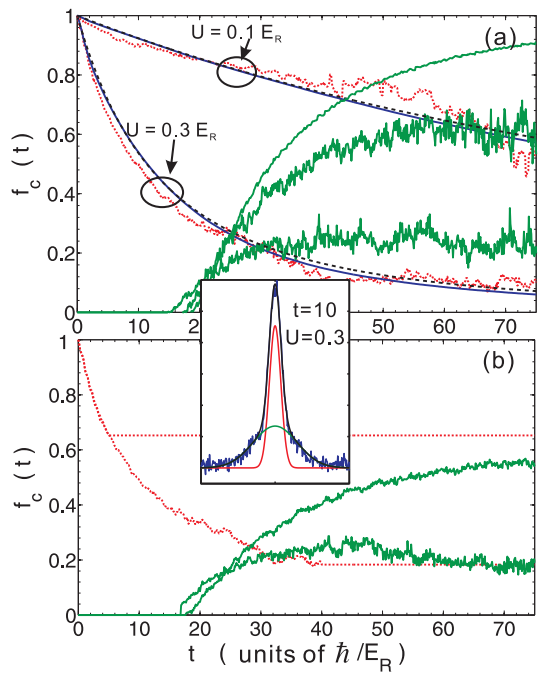}
\caption{(Color online) 
(a) Decay of the condensate fraction, $f_c(t)$ for different interaction energies, $U/E_R=0.1$ and $0.3$. The dotted red and solid blue lines are obtained from the bi-modal fitting [\oncite{N_c}] in momentum distribution and  wavefunction overlapping [\oncite{overlapping}], respectively, while the dashed black lines is fitted by Eq. (\ref{f_c}).
The three raising green lines are the pseudo-condensate-fraction, $\tilde{f}_c(t)$, for $U/E_R=0$, 0.1, and 0.3 (in the order from top to bottom). Inset is the momentum distribution with the bi-modal fitting cures. (b) Same as the case in (a) for $U/E_R=0.3$, but the interaction is turned off at $t_\ast=5\hbar/E_R$ (upper line) and $t_\ast=40\hbar/E_R$ (lower line), respectively. 
}
\label{decay2}
\end{figure}

At the first glance, the similar real space distributions of the GPE and TWA in Fig. \ref{main}(a1)-(a4) seem to suggest that the quantum fluctuation is negligible. However, from the momentum distribution shown in Fig. \ref{main}(b1)-(b4), one can see that the TWA result deviates from the non-interacting or the GPE results seriously: the condensate particles in the central peak are depleted and more non-condensate particles are generated with finite momenta, filling up the valleys of the initial momentum distribution. Note that the GPE result also has valleys after a longer time expansion, which simply results from the finite size effect of the initial wavefunction. In order to quantitatively investigate such a dynamical quantum depletion process, here we define the condensate fraction $f_c(t)$ to be the ratio between the number of condensate particles $N_c(t)$ near the zero momentum at time $t$ to its initial value, i.e., $f_c(t)\equiv N_c(t)/N_c(0)$.
We use two independent numerical methods to extract $f_c(t)$ from our TWA results: one is from the standard bi-modal fitting of the central peak in the {\it momentum} distribution (see Ref. [\oncite{N_c}] and the inset of Fig. \ref{decay2}), and the other one is by calculating the probability to find particles in the condensate coherent state [\oncite{overlapping}], which is supposed to follow the GPE result if no quantum fluctuation is included.
In Fig. \ref{decay2}(a) we show the calculated results obtained from these two methods in the dotted red and solid blue lines, respectively, and find a very good agreement: the condensate fraction decreases monotonically during the TOF expansion and the decay rate becomes larger for a stronger interaction. This quantum fluctuation effect cannot be obtained from our GPE (mean-field) results at all. We also note that when this method will fail automatically if the initial state is a normal or Mott state so that $N_c(0)=0$. 

For comparison, now we further calculate the pseudo-condensate-fraction, $\tlf_c(t)$, which is extracted from the bi-modal fitting in the {\it spatial} density distribution (see Ref. [\oncite{N_c}]), as the experimental convention.
In Fig. \ref{decay2}(a), we show the calculated $\tlf_c(t)$ (raising green lines) for three different interaction strengths: it grows monotonically in time and eventually saturates at a value, $\tlf_c(\infty)$, which is equal to one for a free expansion but smaller than one with a finite interaction.
This result indicates that if the expansion is non-interacting, one can correctly measure the true condensate fraction after a long time expansion ($\tlf_c(\infty)=1$).
On the other hand, when the interaction is considered, $\tlf_c(t)$ saturates at a value  $\tlf_c(\infty)<1$, which cannot represent the true condensate fraction inside the optical lattice. Since the results shown in Fig. \ref{decay2}(a) are all calculated from the same initial condition, we conclude that the spatial density distribution (i.e., the TOF image) cannot give a good measurement of the condensate fraction during the TOF. 

Although the time-dependence of the true condensate fraction, $f_c(t)$, and the pseudo-condensate fraction, $\tlf_c(t)$, are quiet different, their close relationship can be revealed by switching off the particle interaction immediately at time $t_\ast$ during the expansion.
As shown in Fig. \ref{decay2}(b),  $\tlf_c(t)$ evolves eventually to a value set by the true condensate fraction at $t_\ast$, i.e., $\tlf_c(\infty)=f_c(t_\ast)$, no matter its value is larger or smaller than $f_c(t_\ast)$. This is simply because, after $t_\ast$, particles expand freely in space and therefore the momentum distribution at $t_\ast$ is transformed to the spatial distribution.
This also confirms the decay of the condensate fraction.
Note that, without turning off the interaction during the expansion, the true condensate fraction, $f_c(t)$, is hidden in the momentum distribution and therefore cannot be directly observed in the TOF image.
However, turning off the inter-particle interaction can also reduce the expansion speed, making it more difficult to reach the desired saturated value, $\tlf_c(\infty)$. 


The quantum depletion during the TOF can be viewed as a radiative process in the momentum space, where a pair of condensate particles are scattered out of the condensate and transfer their interaction energy toward the kinetic energy.
Such dynamical process is governed by  $H_I'=\frac{g_D}{2\Omega_D}\sum_{\bfq\neq 0}\left[b^\dagger_{\bfq} b^\dagger_{-\bfq} b^{}_0 b^{}_0+{\rm h.c.}\right]$, where $g_D$ is the interaction strength in the $D$-dimensional free space, $b_\bfq$ is the boson operator, and $\Omega_D$ is the system volume.
Here the scattering of particles in the Bragg peaks are neglected for simplicity. Denoting the many-body wavefunction at time $t$ by $|\Psi_0(t)\rangle$, the many-body wavefunction at time $t+dt$ ($dt\to 0^+$) can be described by the following form: $|\Psi_0(t+dt)\rangle=B_0(dt)|\Psi_0(t)\rangle+\sum_{\bfq} B_\bfq(dt) |\Psi_\bfq(t)\rangle$, where $|\Psi_\bfq(t)\rangle\equiv N_c(t)^{-1}b^\dagger_\bfq b^\dagger_{-\bfq}b^{}_0b^{}_0|\Psi_0(t)\rangle$ and $B_\bfq(dt)$ is the associate coefficient.
The non-condensate particle number, $\langle\Psi_0|b^\dagger_\bfq b^{}_\bfq|\Psi_0\rangle$, has been neglected when compared to $N_c(t)\equiv\langle\Psi_0(t)|b^\dagger_0 b^{}_0|\Psi_0(t)\rangle$ in the normalization constant.
Therefore, the condensate particle number at $t+dt$ becomes $N_c(t+dt)=N_c(t)-2\sum_\bfq |B_\bfq(dt)|^2$.
Giving $B_i(0)=\delta_{i,0}$, we can apply the spirit of Fermi-Golden rule for $dt\to 0$, and obtain the decay rate of the condensate density $n_c(t)\equiv N_c(t)/\Omega_D$ by   
\be
\frac{dn_c}{dt} &=&-\frac{4\pi}{\hbar}
\frac{1}{\Omega_D}\sum_{\bfq\neq 0}\left|\langle \Psi_\bfq|
H_I'|\Psi_0\rangle\right|^2\delta(2\mu_D-\hbar^2\bfq^2/m),
\nonumber\\
&\approx &-\frac{4\pi}{\hbar}\frac{2g_D^2 N_c(t)^2}{2\Omega^2}
\int \frac{d^D\bfq}{(2\pi)^D}
\delta\left(2\mu_D-\hbar^2\bfq^2/m\right),
\nonumber\\
&\approx & -\frac{2C_D}{D} g^{2}_Dn_c(t)^{2}\mu_D^{D/2-1},
\ee
where $C_D\propto \frac{1}{\hbar}\left(\frac{m}{\hbar^2}\right)^{D/2}$ is a constant, and $\mu_D$ is the chemical potential in the $D$-dimensional space.
We note that in general $\mu_D$ depends on the interaction strength, global trapping potential, and dimensionality etc.
For example, in a uniform 3D system, $\mu\sim g_3 n_c$ in the weakly interacting limit, while there is no
condensate particles in a uniform 1D system. For a 1D finite size interacting system here, without loss of generality, we may simply assume $\mu_D=\epsilon_Dn_c$, where $\epsilon_D$ should depend on the interaction strength as well as system size. After integrating over this differential equation, we obtain the condensate fraction, $f_c(t)\equiv n_c(t)/n_c(0)$, as following:
\be
f_c(t)&=&\left[\frac{1}{1+\tilde{C}_D n_c(0)^{D/2}g^2 t}\right]^{2/D},
\label{f_c}
\ee
with a non-universal fitting parameter here $\tilde{C}_D\equiv C_D\epsilon_0^{D/2-1}$ (depending on finite size effect, finite temperature, initial non-condensate particles etc.).
It is important to note that, instead of an exponential decay expected in a regular radiative process, the many-body interaction matrix element, $H_I'$, couples the decay rate to the condensate density, so that it becomes an algebraic decay instead.
If the initial state is not a pure condensate, Eq. (\ref{f_c}) should be still valid if only away from the regime of quantum phase transition or Mott phase.
Using Eq. (\ref{f_c}) to fit our TWA results shown in Fig. \ref{decay2}(a) with a fitting parameter $\tilde{C}_D$, we find a very good agreement to confirm the quantum fluctuation effects on the condensate fraction during the TOF experiment. 

\begin{figure}
\includegraphics[width=7cm]{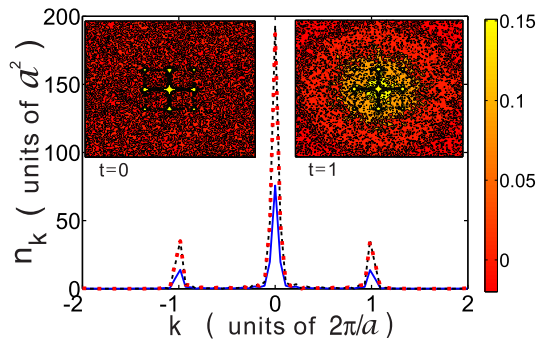}
\caption{(Color online) Momentum distribution of a TOF expansion in 2D space along $y=0$ at $t=\hbar/E_R$. 
The dashed black, dotted red, and solid blue lines are the non-interacting, GPE and TWA results, respectively. (The first two cannot be distinguished here.) The two insets show the momentum distributions for TWA at $t=0$ and $t=\hbar/E_R$, respectively. Here the interaction energy is larger ($U=0.45 E_R$) in order to demonstrate the quantum depletion effects within a short time expansion.}
\label{2D_exp}
\end{figure}
As for higher dimensions, in Fig. \ref{2D_exp}, we also show the momentum distribution of atom clouds expanding in a 2D system from an initial $11\times 11$ lattice sites (total $128\times 128$ mini-cellls).
Due to the limitation of our computational resources, only results of a very short time expansion can be obtained before atoms reach the boundary.
No quantitative information can be extracted yet.
However, by comparing the momentum distributions in the GPE and TWA results, it is easy to find the decay of condensate density even in such a short-time regime due to the quantum fluctuations.

Before concluding, we mark several aspects to discuss: (i) We have also calculated the TOF dynamics of twenty atoms in ten lattice sites (not shown here), and find a stronger decay of $f_c(t)$ for the same interaction strength.
However, results of twenty particles in twenty wells are found almost the same as results of ten particles in ten wells. This shows that the system size is not relevant here, and its dependence of the particle density (rather than total particle number) is also consistent with the nature of quantum fluctuations.
(ii) Since TOF is known as a far-from-equilibrium dynamical process, the integrability of 1D Bose gas should be irrelevant to any of our results here. This is also confirmed by the similar depletion process in 2D systems shown in Fig. \ref{2D_exp}. (iii) From Eq. (\ref{f_c}), it is easy to see that for an expansion in higher dimensional systems, the exponent of  decay becomes smaller, due to the smaller density of states in the long-wavelength limit.
This is therefore also consistent with the general principle that the quantum fluctuation effects become smaller in higher dimensional systems.

In conclusion, we numerically and analytically study the interaction effects for an expanding condensate released from an optical lattice. The condensate fraction has an algebraic decay in time due to the quantum fluctuation effects. Our results provide a deeper understanding to the quantum fluctuation effects during the time-of-flight experiment.
 
We appreciate the discussion with T.-L. Ho, C. Chin, I. Spielman, T. Proto, and G. Modugno. We also thank P.-C. Chen for his kind assistance on the computation.
This work is supported by NSC(Taiwan).



\begin{thebibliography}{99}
\bibitem{N_c}
Our bi-modal fitting process is following: We first set a reasonable momentum $k_0$ and fit the distribution by a Gaussian function for $|k|>k_0$. After substracting this first fitting curve, we fit the residual distribution by a second Gaussian function for $|k|<k_0$. The area of the second fitting result gives $N_c(t)$, which is checked to be independent of $k_0$. Condensate particles at Bragg peaks are neglected. Similar method is applied to extract $\tilde{f}_c(t)$ from the real space distribution.
\end{thebibliography}
\end{document}